\documentclass[12pt,a4paper]{article}
\pdfoutput=1

\usepackage{amsmath,amsfonts,amsbsy,amssymb,enumerate,array}
\usepackage{graphicx,latexsym,verbatim,psfrag}
\usepackage[usenames]{color}
\usepackage[english]{babel}
\usepackage{fancybox}
\usepackage{slashed}

\usepackage{bm} 
\usepackage{booktabs} 
\usepackage{chngpage} 

\usepackage{psfrag}

\usepackage{epsf}
\usepackage{amsthm}
\usepackage{xfrac}
\usepackage{tikz}
\usepackage[utf8]{inputenc}
\usepackage{cancel}

\usepackage[nosort]{cite}   
\usepackage{mciteplus}

\usepackage[colorlinks=true,      linkcolor=blue,      urlcolor=blue,
            filecolor=blue,      citecolor=blue,       pdfstartview=FitH,
						pdfpagemode=UseNone,      bookmarksopen=true]{hyperref}  
\usepackage[all]{hypcap}     

\usetikzlibrary{arrows,automata,positioning,calc,trees,decorations.pathmorphing,decorations.markings}

\def\eq#1{(\ref{#1})}

\renewcommand{\d}{\textrm{d}}
\newcommand{\Tr}{\textrm{Tr}}

\newcommand{\w}{\wedge}

\newcommand{\N}{{\cal{N}}}

\newcommand\varpm{\mathbin{\vcenter{\hbox{%
  \oalign{\hfil$\scriptstyle+$\hfil\cr
          \noalign{\kern-.3ex}
          $\scriptscriptstyle({-})$\cr}%
}}}}

\newcommand\varmp{\mathbin{\vcenter{\hbox{%
   \oalign{\hfil$\scriptstyle-$\hfil\cr
           \noalign{\kern-.3ex}
          $\scriptscriptstyle({+})$\cr}%
}}}}

\newcommand{\YM}{\textrm{\tiny YM}}

  \def\cL{{\cal L}}
 \def\cN{{\cal N}}

\def\be{\begin{equation}}
\def\ee{\end{equation}}
\def\bea{\begin{eqnarray}}
\def\eea{\end{eqnarray}}
\def\bal{\begin{align}}
\def\eal{\end{align}}
\def\nn{\nonumber}

\newcommand{\captionfonts}{\small}

\makeatletter  
\long\def\@makecaption#1#2{%
  \vskip\abovecaptionskip
  \sbox\@tempboxa{{\captionfonts #1: #2}}%
 \ifdim \wd\@tempboxa >\hsize
    {\captionfonts #1: #2\par}
  \else
    \hbox to\hsize{\hfil\box\@tempboxa\hfil}%
  \fi
  \vskip\belowcaptionskip}
\makeatother   

\definecolor{cardinal}{rgb}{0.6,0,0}
\definecolor{darkgreen}{rgb}{0,0.4,0}
\definecolor{golden}{rgb}{0.92, 0.7, 0}
\definecolor{midnight}{rgb}{0, 0, 0.5}
\definecolor{darkblue}{rgb}{0, 0, 0.7}

\begin{document}

\numberwithin{equation}{section}

\begin{flushright}
IPHT-T16/012\\
\end{flushright}

\vspace{23mm}

\begin{adjustwidth}{-15mm}{-15mm} 
 \begin{center}
{\LARGE \textbf{{Loop corrections to the antibrane potential}}}

\vspace{14mm} {\large Iosif Bena, ~Johan Bl{\aa}b{\"a}ck, ~David Turton}\\

\vspace{12mm}

Institut de Physique Th\'eorique, \\
Universit\'e Paris Saclay,\\
CEA, CNRS, F-91191 Gif sur Yvette, France \\

\vspace{10mm} {\small\upshape\ttfamily iosif.bena, johan.blaback, david.turton @ cea.fr} \\

 \vspace{30mm}


\textbf{Abstract}
\end{center}
\end{adjustwidth}

\vspace{3mm}

\begin{adjustwidth}{6.5mm}{6.5mm} 
Antibranes provide some of the most generic ways to uplift Anti-de Sitter flux compactifications to de Sitter, and there is a growing body of evidence that antibranes placed in long warped throats such as the Klebanov-Strassler warped deformed conifold solution have a brane-brane-repelling tachyon. This tachyon was first found in the regime of parameters in which the backreaction of the antibranes is large, and its existence was inferred from a highly nontrivial cancellation of certain terms in the inter-brane potential.
We use a brane effective action approach, similar to that proposed by Michel, Mintun, Polchinski, Puhm and Saad in arXiv:1412.5702, to analyze antibranes in Klebanov-Strassler when their backreaction is small, and find a regime of parameters where all perturbative contributions to the action can be computed explicitly. We find that the cancellation found at strong coupling is also present in the weak-coupling regime, and we establish its existence to all loops.
Our calculation indicates that the spectrum of the antibrane worldvolume theory is not gapped, and may generically have a tachyon. Hence uplifting mechanisms involving antibranes remain questionable even when backreaction is small.

\end{adjustwidth}

\thispagestyle{empty}
\newpage

\baselineskip=12.3pt
\parskip=2pt

\tableofcontents

\baselineskip=15pt
\parskip=3pt

\section{Introduction}

The physics of antibranes in backgrounds with charges dissolved in fluxes has been the subject of intense study in recent years. This physics can be studied in three regimes of the parameter $g_s {N}$, which determines the strength of the backreaction of the antibranes. The first regime is when the backreaction dominates in a region which is large compared to the string scale ($g_s {N} \gg 1$).
The second is when the backreaction of the antibranes is small in any region where supergravity can be trusted  ($g_s {N} \ll 1$), however one does not truncate to leading order in $g_s {N}$.
The third regime is when one truncates to leading order in $g_s{N} \ll 1$; this is sometimes referred to as working in the $g_s {N} \to 0$ limit.

The most commonly used systems for studying the physics of antibranes have D3, M2, or D6 charges dissolved in flux, such as the Klebanov-Strassler (KS) warped deformed conifold background~\cite{Klebanov:2000hb}, the Cvetic-Gibbons-Lu-Pope (CGLP) warped Stenzel background~\cite{Cvetic:2000db}, and the Janssen-Meessen-Ort\'{i}n solution with finite Romans mass~\cite{Janssen:1999sa}.
The most precise calculations of the physics of antibranes have been done in the first (large backreaction) regime. In this regime it was shown that antibrane solutions have a singularity \cite{Bena:2009xk,Bena:2010gs,Dymarsky:2011pm,Bena:2011wh,Blaback:2011pn,Bena:2012bk,Gautason:2013zw,Giecold:2013pza} which cannot be resolved by brane polarization when the antibranes are smeared and their worldvolume is flat \cite{Bena:2012tx,Bena:2012vz,Bena:2014bxa}, and moreover cannot be cloaked by a black hole horizon~\cite{Bena:2012ek,Bena:2013hr,Buchel:2013dla,Blaback:2014tfa}.\footnote{Other antibrane singularities such as those corresponding to antibranes with non-flat (Anti-de Sitter) worldvolumes \cite{Apruzzi:2013yva,Junghans:2014wda} can be resolved by brane polarization. There are also antibrane singularities that can be cloaked with a horizon \cite{Hartnett:2015oda}.
However, the physics of these antibranes is very different from that of antibranes in long warped throats, and hence these calculations have limited relevance for the viability of anti-D3 brane uplifting constructions~\cite{Kachru:2003aw}. For example, the antibranes with non-flat worldvolume only polarize when their worldvolume cosmological constant is parametrically large~\cite{Junghans:2014wda}. Similarly, the cloaked solutions of \cite{Hartnett:2015oda} have a very non-generic type of transverse fluxes which allow them to evade the blackening no-go theorem of \cite{Blaback:2014tfa}, but it is hard to see how antibranes in generic transverse fluxes could do the same \cite{Cohen-Maldonado:2015ssa}.}
Furthermore, if one studies the polarization potential of anti-D$3$ branes localized at a point on the three-sphere at the bottom of the KS solution, one finds that these anti-D$3$ branes generically have a brane-brane-repelling tachyon on their worldvolume \cite{Bena:2014jaa}, which may be responsible for the fact that their singularity cannot be cloaked by an event horizon.

The third regime of parameters described above corresponds to discarding all physics beyond leading order in $g_s{N} \ll 1$.
In this regime, one can study probe anti-D3 branes in the solution S-dual to the KS geometry. One finds that the probe action describing the polarization of these branes into D5 branes has a metastable minimum~\cite{Kachru:2002gs}; this result has been extrapolated to the original KS regime to argue that anti-D3 branes polarize into NS5 branes and give rise to metastable KS minima~\cite{Kachru:2002gs}.
However, as discussed in Ref.~\cite{Bena:2014jaa}, such polarization can only be reliably described when $g_s {N}\gg~\!\!1$.
Furthermore, as explained in Ref.~\cite{Bena:2006rg}, calculations that ignore subleading effects in $g_s {N}$ can give misleading results about metastable vacua: a brane configuration that appears metastable in the $g_s {N} \to 0 $ limit~\cite{Bena:2006rg,Franco:2006es,Ooguri:2006pj} can in fact correspond to a vacuum of a different theory, and this can only be seen by studying the system at finite $g_s {N}$.\footnote{One can also see this from the fact that the action of the tunneling instanton diverges: when $g_s{N}>0$ the distance between the supersymmetric and non-supersymmetric brane configurations diverges at spatial infinity; in the $g_s {N} \to 0 $ limit, this distance is finite but the tension of the branes diverges, and so the tunneling process cannot take place \cite{Bena:2006rg}.}

Hence, in order to investigate further whether antibranes may or may not be metastable in long warped throats, the only regime amenable to calculations that remains to be explored is the second one, $0 < g_s {N} \ll 1$. In an interesting paper, Michel, Mintun, Polchinski, Puhm, and Saad have argued \cite{Michel:2014lva} that in this regime, the correct way to describe one or several antibranes in a background with positive charge dissolved in the fluxes is to use a so-called ``brane effective action''; this action is obtained by integrating out heavy degrees of freedom to obtain an effective field theory (EFT) of light fields on the brane interacting with supergravity fields~\cite{Damour:1975uj,Goldberger:2001tn}.

The exploration of Ref.~\cite{Michel:2014lva} leaves open the question of whether or not antibranes in the KS solution have a brane-brane-repelling tachyon of the type found in \cite{Bena:2014jaa}. Indeed, upon examining the brane effective action of anti-D3 branes localized at the North Pole of the $S^3$ at the bottom of the KS solution, one can easily see that all the terms of this action must transform in representations of the $SO(6)$ R-symmetry group. For example, the interaction potential between two branes is a combination of an $SO(6)$ singlet and a term transforming in the {$\mathbf{20'}$} \cite{Bena:2015qfa}, that furthermore must be invariant under the $SO(3) \times SO(3)$ symmetry preserved by the background and one of the branes. The absence of a tachyon depends on the exact balance of these terms: if the term in the ${\bf 1}$ is stronger than the one in the ${\bf 20'}$ then there is no tachyon, but, if the term in the ${\bf 20'}$ is stronger than the term in the ${\bf 1}$, there will always exist a tachyon.

The purpose of this paper is to identify a regime of parameters in which the brane effective action describing localized anti-D3 branes in the KS solution can be computed, and to use it to evaluate the inter-brane potential to all orders in perturbation theory. As we will discuss in Section \ref{sec:phys}, the diagrams in the brane effective action approach of Ref.~\cite{Michel:2014lva}
correspond to string diagrams in the limit of massless closed strings. For example, at one loop, the string diagram is an annulus.
In the opposite field-theory limit in which the open strings become light, the same string diagram corresponds to a one-loop diagram in the worldvolume gauge theory of the anti-D$3$ branes. Similarly, the higher-loop diagrams in this theory correspond to limits of string diagrams with more than two boundaries.
Since this limit allows explicit computations to be performed, we work in the low-energy gauge theory on the branes.

We first compute the tree-level action, including the terms that are induced by the supergravity background fields.
Upon placing anti-D$3$ branes in a background with a transverse three-form flux, the fermions on the branes acquire a mass, proportional to the value of the imaginary self-dual (ISD) component of the flux \cite{Grana:2002tu}. The ISD three-form flux also induces a scalar trilinear interaction~\cite{Myers:1999ps,Taylor:1999gq}.
%
In addition, anti-D$3$ branes placed in transverse fluxes will generically also have tree-level scalar masses, that can be obtained by expanding the brane potential to quadratic
order.\footnote{When anti-D$3$ branes are placed in an imaginary anti-self-dual (IASD) background, the fermions are massless. In addition, if the background is of IASD Gra\~{n}a-Polchinski-GKP-type~\cite{Grana:2000jj,Giddings:2001yu}, the antibranes feel no potential when moving in the transverse directions, and hence the six scalars are also massless.}

The tree-level fermion and scalar masses in the action of anti-D$3$ branes placed in the KS solution (or similar supersymmetric ISD backgrounds) have three important properties. Firstly, of the six Hermitian scalars, three are massive with equal masses, while three are massless.
Secondly, of the four Weyl fermions, three are massive and one is massless, and the mass-squared of the three massive fermions is half that of the three massive scalars.
Thirdly, the scalar trilinear and the fermion mass term obey a very simple linear relation, discussed in more detail in Section \ref{ssec:superpot}.

The first property follows from the fact that the functions entering in the KS solution only depend on the radial coordinate $\tau$. This implies that there is a flat potential, and hence no force preventing antibranes from moving along the three directions inside the large $S^3$ at the bottom of the KS solution.
The second property is even more intriguing, and is a key feature of anti-D$3$ branes in KS.
There are several ways to see this;
the most straightforward way would be to match the multiple conventions for these terms and compute them directly. However, we will instead derive this property by computing the potential for a probe anti-D$3$ brane in the KS solution to polarize into a D$5$ brane wrapping the contracting $S^2$ of the warped deformed conifold. This potential does not allow for brane polarization, but one finds that the quadratic term is twice larger than it would be if the polarization potential were a perfect square. This is described in Appendix \ref{sec:D5}. This verifies that the three scalars corresponding to the motion of the brane away from the bottom of the warped deformed conifold have a mass-squared that is twice the would-be supersymmetric value.

These two calculations indicate that the sum of the squares of the tree-level scalar masses and the sum of the squares of the tree-level fermion masses are the same, which agrees with the more general result recently found in \cite{Bena:2015qfa} that this is a property of all D3 branes at equilibrium.
Note however that the traceless part of the scalar mass matrix depends on the features of the geometry near the location of the branes and hence is not determined by the fermion masses.

Having obtained the tree-level brane action, we next compute the field-theory loop corrections.
%
The easiest way to compute these corrections is to observe that the antibrane worldvolume theory has the following structure. Consider $\N=4$ Super-Yang-Mills (SYM) theory, broken to $\N=1^*$ by giving equal masses to the three chiral multiplets. As we will show, the antibrane worldvolume theory is a particular $\N=0^*$ theory originating from this equal-mass $\N=1^*$ theory by the addition of a traceless scalar bilinear term (a B-term) that breaks the remaining supersymmetry but preserves the $SO(3) \times SO(3)$ symmetry.
One can then apply a combination of certain general results on finiteness obtained by
Parkes and West~\cite{Parkes:1982tg,Parkes:1983ib,Parkes:1983nv,Parkes:1984dh}
to find that this theory is finite to all orders in perturbation theory.
Thus the masses of the chiral multiplets and the B-terms receive no perturbative corrections, which implies that the inter-brane potential along the $S^3$ at the bottom of the deformed conifold is exactly zero. This is the main result of our paper.

This result is all the more striking because it agrees exactly with the result obtained when $g_s \overline{N}_{\!3} \gg 1$ by studying backreacted antibranes in the KS solution \cite{Bena:2014jaa}. Indeed, one can compute the Polchinski-Strassler~\cite{Polchinski:2000uf} polarization potential of fully-backreacted anti-D$3$ branes localized in the KS solution and one finds, after a pair of surprising cancellations, that the quadratic piece in the polarization potential along the $S^3$ at the bottom of the warped deformed conifold is exactly zero.\footnote{It is important to emphasize that the underlying reason for this cancellation is not understood, and that the cancellation is not simply a consequence of the symmetry of the solution before the branes are added. In the CGLP solution \cite{Cvetic:2000db}, which has a similar symmetry, the potential between anti-M2 branes moving at the bottom of the throat is repulsive \cite{Bena:2014bxa}.} This in turn indicates that the three scalars that describe the motion of the anti-D$3$ branes at the bottom of the KS solution are massless when $g_s \overline{N}_{\!3} \gg 1$.

The fact that this strong-coupling result agrees with the all-loop perturbation theory result suggests that the masslessness of the three scalars of the worldvolume theory of anti-D$3$ branes in KS is a property that is valid for all values of $g_s \overline{N}_{\!3}$. Of course it is logically possible that as one increases the coupling the potential will rise and fall again, as a smooth function of compact support, but this appears rather implausible.
 In our opinion the fact that, despite the absence of supersymmetry, this potential is exactly zero in the only two regimes of parameters where it has been computed exactly, strongly suggests that this potential will be zero throughout.
 Moreover, this flat direction means that in our regime of parameters, the spectrum is not gapped, and is possibly tachyonic. We will discuss this in detail in Sections  \ref{sec:phys} and \ref{sec:discussion}.

This paper is organized as follows. In Section \ref{sec:Bulk} we review the KS solution, extracting the properties required for our analysis. We derive the bosonic part of the worldvolume theory of anti-D$3$ branes in the KS solution in Section \ref{ssec:wv}, and the fermionic part in Section \ref{ssec:superpot}. We describe how the non-renormalization theorems are applied to our theory in Section \ref{ssec:nonrt}. In Section \ref{sec:phys} we discuss the physical interpretation and the regime of validity of our result. We end with a discussion in Section \ref{sec:discussion}.

\section{From bulk solutions to worldvolume theories}\label{sec:Bulk}

\subsection{The Klebanov-Strassler background}\label{sec:KS}

The purpose of this subsection is to review some properties of the KS background \cite{Klebanov:2000hb}, to extract certain relations that we need for our analysis, and to introduce notation. We will not give a full review of KS; for more details we refer the reader to Ref.~\cite{Herzog:2001xk}, which agrees with most of our conventions.

The KS background is a supersymmetric, non-compact, Gra\~na-Polchinski-GKP-type~\cite{Grana:2000jj,Giddings:2001yu} solution. By this we mean the following. The $G_3 \equiv F_3 - i e^{-\phi}H_3$ flux has (2,1) complexity, the dilaton $e^{\phi}=g_s$ is constant, and the  ten-dimensional (string-frame) metric is a warped product of $\mathbb{R}^{3,1}$ with a Calabi-Yau base:
\begin{equation}
\d s_{10}^2 = e^{2A} \d \tilde{s}^2_4 + e^{-2A} \d \tilde{s}^2_6\,.
\end{equation}
As this expression suggests, the metrics which have a tilde are unwarped. The $C_4$ potential takes the form\footnote{We will use the notation that  $F_5 = \star_{10} \d C_4$, or equivalently that $F_5$ is the internal part of the self-dual five-form field strength $\mathcal{F}_5 = (1+\star_{10})F_5$.}
\begin{equation}
C_4 = \tilde{\star}_4 \alpha\,
\end{equation}
where the function $\alpha$ is related to the warp factor $e^{4A}$ by\footnote{Note that we are using the string-frame metric throughout this paper. This means that there is an extra factor of the (constant) dilaton in this expression compared to the corresponding  formula in \cite{Giddings:2001yu}, written using the Einstein-frame metric.}
\begin{equation}\label{eq:GKPrel}
\alpha = e^{-\phi}\left(e^{4A} - \alpha_0\right) ,
\end{equation}
where $\alpha_0$ is a constant that we gauge fix by requiring that $\alpha(\tau=0)=0$.

In the KS solution, the Calabi-Yau base is a deformed conifold with topology $\mathbb{R}^+ \times S^2 \times S^3$, and all the functions that determine the solution, such as $\alpha$ and $A$ above, depend only on the radial direction of the deformed conifold, which is commonly denoted as $\tau$ and parameterizes the $\mathbb{R}^{+}$.

This solution has four supercharges, which are compatible with those of D3 branes. Hence, probe anti-D$3$ branes experience a potential that forces them to the bottom of the throat,
\begin{equation}
V \sim \mu_3 e^{-\phi} \sqrt{-g_4} + \mu_3 \alpha = 2\mu_3 e^{-\phi}\left( \frac{1}{2}e^{4A_0} + \partial_\tau^2 e^{4A}|_{0} \tau^2 + \ldots \right)\,.
\end{equation}

In this paper we consider anti-D$3$ branes localized at the bottom of the deformed conifold, and hence we employ a local $\mathbb{R}^6$ coordinate system \cite{Bena:2014jaa} parameterizing the neighborhood of the branes,
\begin{equation}
\d s^2_{10} = e^{2A} \eta_{\mu\nu} \d x^\mu \d x^\nu + e^{-2A} \tilde{g}_{mn} \d x^m \d x^n\,,
\end{equation}
where the internal metric is given by $\tilde{g}_{mn} = b \delta_{mn}$ for some constant $b$ that depends on  conventions.

In such a local coordinate system, the local expansion in terms of $\tau$ is now an expansion in $x^{7,8,9}$ -- the local coordinates of the $\mathbb{R}^+ \times S^2$. This means that the $\tau^2$ term of the anti-D$3$ brane potential will have the form
\begin{equation}
\begin{split}
\tau^2 &\propto (x_7^2 + x_8^2 + x_9^2)\,,
\end{split}
\end{equation}
which, as mentioned in the Introduction, can be decomposed as the sum of a quadratic term transforming in the {\bf 1} (singlet) of the SO(6) R-symmetry and a term transforming in the  $\mathbf{20'}$ traceless symmetric representation,
\begin{equation}\label{flat-bottom}
\begin{split}
\tau^2 &\propto (x_7^2 + x_8^2 + x_9^2)\\
&= \frac{1}{2} (x_4^2 + x_5^2 + x_6^2 + x_7^2 + x_8^2 + x_9^2) - \frac{1}{2} (x_4^2 + x_5^2 + x_6^2 - x_7^2 - x_8^2 - x_9^2)\,.
\end{split}
\end{equation}
For later convenience we introduce complex coordinates
\begin{equation}\label{eq:cplxcoord}
z_i \equiv \frac{1}{\sqrt{2}}\left( x_{i+3} + i x_{i+6}\right) ,
\end{equation}
with $i=1,2,3$, in terms of which we have
\begin{eqnarray}
\frac{1}{2} (x_4^2 + x_5^2 + x_6^2 + x_7^2 + x_8^2 + x_9^2) &=& z_1 \bar{z}_1 + z_2 \bar{z}_2 + z_3 \bar{z}_3 \,, \label{eq:1and20-1}\\
\frac{1}{2} (x_4^2 + x_5^2 + x_6^2 - x_7^2 - x_8^2 - x_9^2) &=& \frac{1}{2} \left( z_1^2 + z_2^2 + z_3^2  \right) + \textrm{h.c.}
 \label{eq:1and20-20}
\end{eqnarray}

To derive the supersymmetric term on the anti-D$3$ brane worldvolume, we will make use of the $F_5$ Bianchi identity
\begin{equation}\label{eq:bianchi}
\d F_5 = H_3 \w F_3 \,,
\end{equation}
written here for a source-less background, such as the KS solution. Expanding around the bottom of the deformed conifold, and using the fact that $e^{-4A}$ has no linear term in this expansion for KS, we can write the LHS~of Eq.\;\eq{eq:bianchi} as
\begin{equation}
\d F_5 = - e^{-8A_0} \d \tilde{\star}_6 \d \alpha |_{0}\,,
\end{equation}
using the metric and the $C_4$ potential. Hence
\begin{equation}
\d F_5 = e^{-8A_0} b^{-1} \delta^{mn} \partial_m \partial_n \alpha|_{0}\, \tilde{\star}_6 1\,,
\end{equation}
in real coordinates, or in the complex coordinates introduced before
\begin{equation}
\d F_5 = 2 e^{-8A_0} b^{-1} e^{-\phi}\delta^{i\bar{\imath}}(\partial_i \partial_{\bar{\imath}} e^{4A} )|_{0} \tilde{\star}_6 1\,,
\end{equation}
where we also used the relation (\ref{eq:GKPrel}).

To evaluate the RHS~of Eq.\;(\ref{eq:bianchi}) we rely on the imaginary self duality of the  KS $G_3$ flux, which implies that the NSNS and RR fluxes are related by $H_3 = e^{\phi} \star_6 F_3$. Together with $F_3 = \tfrac{1}{2}(G_3 + \bar{G}_3)$, this gives
\begin{equation}
H_3 \w F_3 = \frac{1}{3!} \frac{1}{4} e^{\phi} \left( (G_3 + \bar{G}_3)_{mnp}(G_3 + \bar{G}_3)^{mnp} \right) \star_6 \!1\,,
\end{equation}
where indices are contracted using the warped metric. Transforming to complex coordinates and using the fact that $G_3$ is purely (2,1), only one type of contraction is non-vanishing,
\begin{equation}
(G_3)_{mnp}(\bar{G}_3)^{mnp} = 3 (G_3)_{i j \bar k} (\bar{G}_3)^{i j \bar k}\,.
\end{equation}
Thus we obtain
\begin{equation}
H_3 \w F_3 = \frac{3}{2} e^{\phi} b^{-3} |G_3|^2 \, \tilde{\star}_6 1\,,
\end{equation}
where $(G_3)_{ij\bar{k}} = |G_3| \epsilon_{ij\bar{k}}$ and $|G_3|$ is real. Since $A$ a function of $\tau$ only, we can now write Eq.\;(\ref{eq:bianchi}) as
\begin{equation}\label{eq:bianchiFinal}
\partial_i \partial_{\bar\jmath}e^{4A}|_{0} = \frac{1}{4} g_s^2 b^{-2} e^{8A_0} |G_3|^2 \delta_{i \bar\jmath}\,.
\end{equation}

In due course we shall also make use of the other quadratic terms in the Taylor expansion of $e^{4A}$.
Again using the fact that $A$ is a function of $\tau$ only, we can use Eqs.\;(\ref{flat-bottom})--(\ref{eq:1and20-20}) to find that these terms are given by
\begin{equation}
\label{complex-20}
\partial_i \partial_{j}e^{4A}|_{0}  = - \frac{1}{4} g_s^2 b^{-2} e^{8A_0} |G_3|^2 \delta_{i j}\,.
\end{equation}
These terms transform in the {$\mathbf{20'}$}, unlike the singlet term in (\ref{eq:bianchiFinal}). As one can see from Eqs.\;(\ref{flat-bottom})--(\ref{eq:1and20-20}), the contributions to the potential along the $S^3$ directions coming from the terms in the $\mathbf{1}$ and the $\mathbf{20'}$ are equal and opposite, while along the $\mathbb{R}^+ \times S^2$ directions they add.

\subsection{The bosonic terms in the worldvolume theory}\label{ssec:wv}

We now derive the worldvolume gauge theory of anti-D3 branes at the bottom of the KS throat. We start with the bosonic terms in the action, and we compute the fermionic terms in the next subsection.

The worldvolume gauge field will not play an important role, so for ease of presentation we shall suppress it in what follows, with the understanding that the full action contains the usual gauge kinetic terms and covariant derivative couplings.
In addition, since the $U(1)$ sector of the gauge theory is free, we focus on the $SU(\overline{N}_{\!3})$ sector.
With these points understood, the bosonic part of the theory is given by the DBI and WZ parts of the brane Lagrangian\footnote{The Lagrangian is written with explicit signs corresponding to an antibrane in our conventions. We use the generic term \emph{brane} to refer to both anti-D$p$ and D$p$ branes, where explicit signs and dimensions determine the details.}
\begin{equation}
\begin{split}
\cL_{\textrm{DBI}} &= - \mu_3 \Tr \left\{ e^{-\phi}\sqrt{-\det(P[M_{ab}])\det(Q^{m}_{\phantom{m}n})} \right\}\,,\\
\cL_{\textrm{WZ}} &= - \mu_{3} \left. \Tr\left\{P[e^{i \lambda \imath^2_{\varphi}}C \w e^{B_2}]\right\}\right|_{0123}\,.
\end{split}
\end{equation}
 We work with the string-frame metric. The indices $a, b, \ldots$ are ten-dimensional, the indices $\mu, \nu, \ldots$ are four-dimensional worldvolume indices parallel to the brane, and $m, n, \ldots$ are six-dimensional indices transverse to the brane. The tensors in the above expression are defined as follows:
\begin{equation}
\begin{split}
P[M_{ab}] &\equiv E_{\mu\nu} + E_{\mu m}(Q^{-1}-\delta)^{mn} E_{n\nu}\,,\\
Q^{m}_{\phantom{m}n} &\equiv \delta^{m}_{\phantom{m}n} + i \lambda [\varphi^m,\varphi^p]E_{pn}\,,\\
E_{ab} &\equiv g_{ab} - B_{ab}\,,\\
C &\equiv \sum C_n\,,
\end{split}
\end{equation}
where $\lambda \equiv 2\pi \ell_s^2$ and the D3 brane charge is $\mu_3 = 2\pi/(2\pi \ell_s)^4$,  where $\ell_s$ is the string length.\footnote{Our conventions are described in Appendix \ref{app:conv}.}
The Hermitian scalars $\varphi^m$ transform in the adjoint of $SU(\overline{N}_{\!3})$ and parameterize the brane positions. We also choose a gauge for $B_2$ such that $B_2|_0 = 0$.


Expanding in powers of $\lambda$, the DBI Lagrangian becomes
\begin{eqnarray}
\cL_{\textrm{DBI}} &=& - \mu_3 \Tr \left\{ e^{-\phi}\sqrt{-\det(P[M_{ab}])\det(Q^{m}_{\phantom{m}n})} \right\} \cr
&=& \mu_3 \lambda^2 \left( - \frac{1}{2} e^{-\phi} \tilde{g}_{mn} \Tr\{\partial_\mu \varphi^m \partial_\nu \varphi^n\} \eta^{\mu\nu} - \frac{1}{2} e^{-\phi} \partial_m \partial_n e^{4A}|_{0} \Tr \left\{ \varphi^m \varphi^n\right\} \right.\\
&& {}  \left. -\frac{i}{3} e^{4A_0} e^{-\phi}H_{mnp}\Tr\{\varphi^m \varphi^n \varphi^p\} + \frac{1}{4} e^{-\phi} \tilde{g}_{mq}\tilde{g}_{np} \Tr \left\{ [\varphi^m,\varphi^n][\varphi^q,\varphi^p]\right\}\right) + \ldots  \nonumber
\end{eqnarray}
where as usual the $H_3$ coupling arises from the Taylor expansion of $B_2$, and where
we have dropped the constant term proportional to $e^{4A_{0}}$.

Similarly, expanding the WZ Lagrangian gives
\bea
\cL_{\textrm{WZ}} &=& - \mu_{3} \left. \Tr\left\{P[e^{i \lambda \imath^2_{\varphi}}C \w e^{B_2}]\right\}\right|_{0123}\\
&=&{} \mu_3 \lambda^2 \left( -\frac{1}{2} e^{-\phi} \partial_m \partial_n e^{4A} |_{0} \Tr\{\varphi^m \varphi^n\} -\frac{i}{3} e^{4A_0}\Tr\{ \varphi^m \varphi^n \varphi^p\} (\star_6 F)_{mnp} \right) + \ldots  \nn
\eea
where we have used the fact mentioned below Eq.\;(\ref{eq:GKPrel}) that in our gauge we have $\alpha(\tau=0)=0$ and, hence, at the location of the branes $ \d C_6=F_7 = - \star_{10} F_3$.

The two trilinear terms can be combined into an expression involving a particular combination of $G_3$ and its complex conjugate,
\begin{equation}
\cL_{\rm \,tri} = \frac{\lambda^2 \mu_3}{3}e^{4A_0}(G_3 - \bar{G}_3)_{mnp} \Tr\{\varphi^m \varphi^n \varphi^p\}\,.
\end{equation}
It is convenient to introduce complex coordinates that we shall use from now on,
\begin{equation}
\phi^i = \frac{1}{\sqrt{2}} \left( \varphi^{i+3} + i \varphi^{i+6} \right)\,.
\end{equation}
In these complex coordinates the full Lagrangian, $\cL = \cL_{\textrm{DBI}} + \cL_{\textrm{WZ}}$, becomes
\begin{equation}
\begin{split}
\cL &= \mu_3 \lambda^2 \Bigg[ - e^{-\phi} \tilde{g}_{i\bar{\jmath}} \Tr\{\partial_\mu \phi^i \partial_\nu \bar{\phi}^{\jmath}\} \eta^{\mu\nu}\\
&\qquad\qquad - 2 e^{-\phi} \left( \partial_i \partial_{\bar{\jmath}} e^{4A}|_{0} \Tr\{\phi^{i}\bar{\phi}^{\bar{\jmath}}\} + \frac{1}{2} \left( \partial_{i} \partial_{j} e^{4A}|_{0} \Tr\{\phi^i \phi^j\} + \textrm{h.c.} \right) \right)\\
&\qquad\qquad +  e^{4A_0}\left( G_{ij\bar{k}} \Tr\{\phi^i \phi^j \bar{\phi}^{\bar{k}}\} + \textrm{h.c.} \right)\\
&\qquad\qquad + \frac{1}{2} e^{-\phi} \tilde{g}_{i \bar{\imath}} \tilde{g}_{j \bar{\jmath}}\Tr \{ [\phi^i,\phi^j][\bar{\phi}^{\bar{\imath}},\bar{\phi}^{\bar{\jmath}}] - [\phi^i,\bar{\phi}^{\bar{\jmath}}][\phi^{j},\bar{\phi}^{\bar{\imath}}]\}\Bigg]+ \ldots
\end{split}
\end{equation}
where a bar on the scalars indicates Hermitian conjugation.

In order to proceed we
make a constant rescaling of the scalars to obtain a canonically-normalized kinetic term. This is achieved by defining
\begin{equation}\label{eq:rescale}
\hat{\phi}^i \equiv \phi^i \left( \mu_3 \lambda^2 b e^{-\phi}\right)^{1/2}\,.
\end{equation}
Having done this, we immediately drop the hat from the rescaled expressions and exclusively use the canonically-normalized scalars from now on.

After applying the relation (\ref{eq:bianchiFinal}) derived from the Bianchi identity in the previous subsection, the Lagrangian becomes
\bea\label{eq:Lbranes}
\cL &=& {} - \Tr\{\partial_\mu \phi^i \partial_\nu \bar{\phi}^{\bar\jmath}\} \eta^{\mu\nu} \delta_{i\bar{\jmath}}\cr
&&{} - \left( \frac{g_s}{\sqrt{2}} e^{4A_0} |G_3| b^{-3/2}\right)^2 \delta_{i\bar{\jmath}} \Tr\{\phi^i \bar{\phi}^{\bar{\jmath}}\} -\left( \partial_{i} \partial_{j} e^{4A}|_{0} \Tr\{\phi^i \phi^j\} + \textrm{h.c.} \right)\cr
&&{} + \left(\sqrt{2} \frac{g_s^{1/2}}{\lambda \sqrt{\mu_3}}\right)
\left(\frac{g_s}{\sqrt{2}} e^{4A_0} |G_3| b^{-3/2}\right)
\left(  \epsilon_{ij\bar k} \Tr\{\phi^i \phi^j \bar{\phi}^{\bar{k}}\} + \mathrm{h.c.} \right)\\
&&{} + \frac{1}{2} \left(\sqrt{2}\frac{g_s^{1/2}}{\lambda\sqrt{\mu_3}}\right)^2 \delta_{i \bar{\imath}} \delta_{j \bar{\jmath}} \Tr \left\{[\phi^i,\phi^j][\bar{\phi}^{\bar{\imath}},\bar{\phi}^{\bar{\jmath}}]\right\} - \frac{1}{2} \frac{g_s}{\lambda^2\mu_3} \Tr\{\left( \delta_{i \bar{\jmath}} [\phi^i,\bar{\phi}^{\bar{\jmath}}]\right)^2\}+ \ldots \nn
\eea
Here we have written the quadrilinear term as the sum of an F-term and a D-term, as we will shortly write part of the Lagrangian in terms of an $\cN=1$ superpotential and, as is well-known, the F-term part of the quadrilinear interaction of $\cN=4$ SYM is contained in the superpotential, while the D-term part is not (see for example~\cite{Maldacena:2003zi}).

We can identify the bilinear and trilinear scalar interactions in the language of soft supersymmetry-breaking. To this end we introduce complex-scalar masses $(m_B^2)_{i\bar{\jmath}}$, B-terms $ b_{ij}$, and the (2,1) trilinear interaction $r_{ij\bar{k}}$ via
\bea\label{eq:Lbranes-2}
\cL_{\rm soft} &=& {} - (m_B^2)_{i\bar{\jmath}} \Tr\{\phi^i \bar{\phi}^{\bar{\jmath}}\}
 +\left( - \frac{1}{2} b_{ij} \Tr\{\phi^i \phi^j\}
  + r_{ij\bar k} \Tr\{\phi^i \phi^j \bar{\phi}^{\bar{k}}\} + \mathrm{h.c.} \right). \qquad
\eea
By matching these terms with Eq.\;\eq{eq:Lbranes}, we find the following bosonic soft supersymmetry-breaking terms:
\begin{equation}
\begin{split}
\textrm{Complex-scalar masses: ~} &  (m_B^2)_{i\bar{\jmath}} \;=\; m_B^2 \delta_{i\bar{\jmath}} \,, \qquad m_B \equiv \left( \frac{g_s}{\sqrt{2}} e^{4A_0} |G_3| b^{-3/2}\right)\,,\\
\textrm{B-terms: ~} & b_{ij} \;=\; - m_B^2 \delta_{ij} \,,\\
\textrm{(2,1) trilinear: ~} & r_{ij\bar{k}} \;=\; \left(\sqrt{2} \frac{g_s^{1/2}}{\lambda \sqrt{\mu_3}}\right)
\left(\frac{g_s}{\sqrt{2}} e^{4A_0} |G_3| b^{-3/2}\right) \epsilon_{ij\bar k}\,.
\end{split}
\label{eq:bosonicsoftterms}
\end{equation}

We note that the tree-level B-terms are real. A priori these terms could have had an imaginary part, which would correspond to off-diagonal elements in the real matrix $\partial_m \partial_n e^{4A}|_{0}$.
We will discuss the imaginary part of the B-terms further in Section \ref{sec:phys}.

\subsection{The fermionic terms in the worldvolume theory
}\label{ssec:superpot}

The fermionic part of the action for D3 branes in transverse RR and NSNS three-form fluxes was first explicitly written down in Ref.~\cite{Grana:2002tu} and further studied in Refs.~\cite{Grana:2003ek,Camara:2003ku,McGuirk:2012sb}.
We now compute the terms in this fermionic action and their precise normalization for antibranes in KS.

The calculation deriving the fermion bilinear \cite{Grana:2002tu} and that determining the scalar trilinear \cite{Myers:1999ps} were performed using different conventions, so it is necessary to fix the relative normalizations of the bosonic and fermionic actions. To fix this overall factor, we now examine the form of the fermionic bilinear terms.

Recall that upon writing $\cN=4$ SYM in an $\cN=1$ superfield formalism, one obtains three chiral multiplets.
The KS background has only $(2,1)$ primitive three-form flux, so the only additional term in the fermion action is a mass for the three Weyl fermions in the chiral multiplets, and the gaugino remains massless~\cite{Grana:2002tu}.
We write the (2,1) primitive three-form flux in the form
\bea
S_{\bar \imath \bar \jmath} = \frac12 \left(
\epsilon_{\bar \imath}{}^{kl} G_{kl\bar \jmath}
+\epsilon_{\bar \jmath}{}^{kl} G_{kl\bar \imath}
\right),
\eea
where indices have been raised with $\delta^{i\bar \imath}$. Then the fermion mass matrix is proportional to the complex conjugate of $S$~\cite{Grana:2002tu},
\bea
m_{ij}^{F} &\propto& \left(\bar S\right)_{ij} \,.
\eea
By expanding the $G_3$  flux at the location of the branes, one finds a diagonal fermion mass matrix with equal entries,
\bea
S_{\bar \imath \bar \jmath} = S \, \delta_{\bar \imath \bar \jmath}
\qquad \Rightarrow \qquad
m_{ij}^{F} = m_F \, \delta_{ij} \,.
\eea

Next, we observe that the three massive Hermitian scalars have a mass-squared which is twice the one that they would have if supersymmetry had been preserved.
This can be seen from computing the polarization potential of a D$5$ brane carrying anti-D$3$ brane charge and wrapping the shrinking $S^2$ of the KS background. As outlined in Appendix \ref{sec:D5}, the D$5$ brane polarization potential does not have a finite-radius minimum. However, when the term originating from the mass is halved, the polarization potential becomes a perfect square and has a supersymmetric polarization minimum. Hence the supersymmetric mass-squared is half the mass-squared of the three scalars corresponding to motion away from the tip.
From the argument made around Eq.\;(\ref{flat-bottom}) we see that the supersymmetric mass is the mass of the complex scalars, given in Eq.\;\eq{eq:bosonicsoftterms}. Thus the mass of the Weyl fermions and the complex scalars are equal,
\bea \label{eq:bosonfermionmass}
m_F ~=~ m_B ~=~ \frac{g_s}{\sqrt{2}} e^{4A_0} |G_3| b^{-3/2} \,.
\eea
This determines the correct normalization of the fermionic terms in the Lagrangian, and implies that the sum of the squares of the tree-level scalar masses and the sum of the squares of the tree-level fermion masses are the same. This fact agrees with the more general result recently found in Ref.~\cite{Bena:2015qfa} that this is a property of all D3 branes at equilibrium in warped compactifications.

Having established this relation between fermion and boson masses, we now observe that the  $(2,1)$ scalar trilinear coupling is proportional to the mass of the fermions. This fact, combined with the equality of fermion and boson masses in the absence of the B-term, allows us to temporarily put aside the B-term (and the D-term quadrilinear interaction), and write the remainder of the Lagrangian in terms of $\cN=1$ superfields. Similar observations have been made in Refs.~\cite{Camara:2003ku,McGuirk:2012sb}.

The Lagrangian for these terms can then be written as
\begin{equation}
\cL_{\rm susy} = \Tr \left\{ \bar{\Phi} \Phi \big|_{\theta^2\bar\theta^2}\right\} + \left( W(\Phi)\big|_{\theta^2} + \mathrm{h.c.} \right)\,,
\end{equation}
where $W(\Phi)$ is the superpotential
\begin{equation}
W(\Phi) =   \frac{1}{2}m^F_{ij} \Tr\left\{\Phi^i\Phi^j\right\} + \frac{c \, g_{\YM}}{3} \epsilon_{ijk} \Tr\left\{ \Phi^i\Phi^j\Phi^k  \right\}\,
\end{equation}
where $c$ is a numerical parameter that depends on conventions.
The supersymmetry is broken from $\cN = 4 \to \cN = 1^{\star}$ by the three equal masses of the chiral multiplets. The $\Phi^i$ are the chiral multiplet superfields, written in component fields as
\begin{equation}
\Phi^i = \phi^i + \sqrt{2}\theta \psi^i + \theta^2 F^i\,.
\end{equation}
Upon eliminating the auxiliary fields $F^i$, the Lagrangian becomes
\begin{equation}\label{eq:Lspot}
\begin{split}
\cL_{\rm susy}  &= - \Tr\{\partial_\mu \phi^i \partial_\nu \bar{\phi}^{\bar{\jmath}}\} \delta_{i\bar{\jmath}} \eta^{\mu\nu} + i \Tr\{ (\partial_\mu \bar\psi_i) \bar\sigma^{\mu}\psi_i\}\\
&\quad - \left[(m^F)(\bar{m}^F)\right]_{i\bar\imath} \Tr\{\phi^i \bar{\phi}^{\bar{\imath}}\} - \frac{1}{2} \left( m^F_{ij} \Tr\{\psi^i \psi^j\} + \textrm{h.c.}\right)\\
&\quad
+ \left( c \, g_{\YM} \, \bar{m}^F_{\bar{k}\bar{l}} \, \epsilon_{ij}{}^{\bar{l}} \Tr\{\phi^i\phi^j \bar{\phi}^{\bar{k}}\} + c \, g_{\YM} \, \epsilon_{ijk} \Tr\{ \psi^i \psi^j \phi^k\} + \textrm{h.c.}\right)\\
&\quad + \frac{1}{2} \left| c \, g_{\YM} \right|^2 \delta_{i \bar{\imath}} \delta_{j \bar{\jmath}} \Tr \left\{[\phi^i,\phi^j][\bar{\phi}^{\bar{\imath}},\bar{\phi}^{\bar{\jmath}}]\right\}\,,
\end{split}
\end{equation}
where again indices are contracted with $\delta_{i\bar{\jmath}}$ or its inverse.
We can now read off\footnote{There is a redundancy in conventions in how one exactly chooses the value of the constant $c$, and how one relates $g_{\YM}$ to $g_s$; for our purposes we will not need to fix this redundancy. }
\begin{equation} \label{eq:gymgs}
c \, g_{\YM} = \sqrt{2} \frac{g_s^{1/2}}{\lambda \sqrt{\mu_3}} \,.
\end{equation}
Combining Eqs.~\eq{eq:bosonicsoftterms}, \eq{eq:bosonfermionmass} and \eq{eq:gymgs}, we observe that the relation between fermion masses and scalar trilinear couplings takes the explicit form:
\bea
r_{ij\bar{k}} &=& m_F \left(c \, g_{\YM}  \epsilon_{ij\bar k} \right) \,.
\eea
This simple relation, which is also present in more general theories on D-brane worldvolume, will be a crucial ingredient in our analysis of perturbative corrections, as in general it leads to significant simplifications in beta functions~\cite{Jack:1999ud}.

The supersymmetric Lagrangian (\ref{eq:Lspot}) reproduces the fermionic terms and all terms in the bosonic Lagrangian (\ref{eq:Lbranes}) except for the B-terms and the D-term quadrilinear. Thus we see that the B-terms are the only terms responsible for breaking the $\cN=1$ supersymmetry, and making the theory an $\cN = 0^\star$ theory.
As mentioned earlier, the gauge fields have been suppressed, but can easily be reintroduced.
This concludes the calculation of the anti-D3 brane worldvolume tree-level Lagrangian.

\section{Loop corrections and non-renormalization theorems}\label{ssec:nonrt}

Having derived the tree-level action of the anti-D3 worldvolume gauge theory, we now proceed to investigate quantum corrections. These corrections would generically cause the masses them to run logarithmically with the energy, and this running can be thought of as coming from the backreaction of the anti-D3 branes on the corresponding supergravity fields.

\begin{figure}
\begin{center}\hspace{-10mm}
\begin{tabular}{m{1.9cm}m{0cm}m{1.9cm}m{0cm}m{1.9cm}m{0cm}m{1.9cm}m{0cm}m{1.9cm}m{1cm}}
\includegraphics{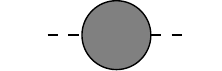}
&=&
\includegraphics{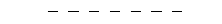}
&$+$&
\includegraphics{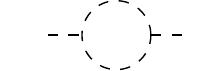}
&$+$&
\includegraphics{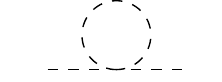}
&$+$&
\includegraphics{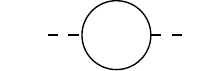}
&$+\ \ldots$
\end{tabular}
\end{center}
\caption{Field-theory loop corrections to the scalar mass, involving the trilinear and quadrilinear scalar couplings and the Yukawa couplings.\label{fig:oneloop}
\vspace{5mm}
}
\end{figure}

\begin{figure}[t!]
\begin{center}
\begin{tabular}{m{.4cm}m{3.02cm}m{.3cm}m{3.05cm}m{.35cm}m{2.8cm}m{.4cm}m{.4cm}}
&
\includegraphics{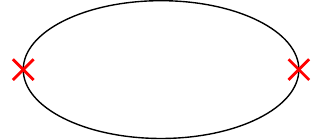}
& $+$ &
\includegraphics{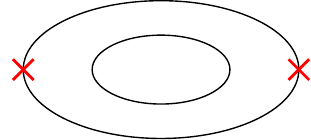}
& $+$ &
\includegraphics{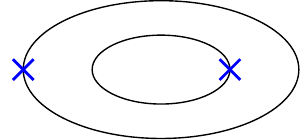}
& $+$ &
\end{tabular}
\begin{tabular}{m{.4cm}m{3.1cm}m{.35cm}m{2.8cm}m{.4cm}m{.4cm}}
 + &
\includegraphics{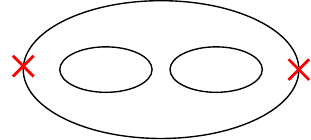}
&+&
\includegraphics{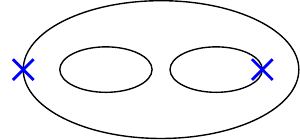}
& + & $\ldots$
\end{tabular}
\end{center}
\caption{Open-string loop expansion. Crosses represents open string vertex operators; red is used for planar diagrams, and blue for non-planar diagrams. \label{fig:openstring}}
\vspace{3mm}
\end{figure}

The worldvolume gauge theory of a stack of $\overline{N}_{\!3}$ coincident anti-D$3$ branes is a $U(\overline{N}_{\!3}) = SU(\overline{N}_{\!3})\times U(1)$ theory. All of the interaction terms derived above, except the mass terms and the B-terms, are anti-symmetrized and hence, as usual, the $U(1)$ sector is free and decouples. In the $SU(\overline{N}_{\!3})$ sector, the diagrams that provide the corrections to the masses of the scalars are summarized schematically in Figure \ref{fig:oneloop}.
These diagrams are the usual field-theory limit of open-string diagrams (see Figure \ref{fig:openstring}). These diagrams come with a factor of $g_s \overline{N}_{\!3}$ for each additional boundary, and hence we are in a regime of perturbative control when $g_s \overline{N}_{\!3} \ll 1$. Thus, naively, one would expect a one-loop correction to the scalar mass of order
\begin{equation}
\frac{\delta_{(1)} (m_B^2)}{(m_B^2)_{\textrm{tree}}} \propto g_s \overline{N}_{\!3}\,.
\end{equation}
%

\vspace{15mm}

In the previous section we derived the structure of how supersymmetry is broken at tree level on anti-D$3$ branes in KS,
\begin{equation}
\underbrace{W_4 = \frac{c \, g_{\YM}}{3} \epsilon_{ijk} \Tr\left( \Phi^i\Phi^j\Phi^k  \right)}_{\cN = 4} \to \underbrace{W_1 = W_4 + \frac{1}{2}m^F_{ij} \Tr\left(\Phi^i\Phi^j\right)}_{\cN = 1^{\star}} \to \underbrace{W_1\ \&\ \textrm{B-terms}}_{\cN = 0^{\star}}\,.
\end{equation}
The purpose of this section is to describe how this type of supersymmetry breaking affects the running of the masses and couplings. Since along the $S^3$ directions there is a perfect cancellation between the supersymmetric scalar mass terms and the real part of the supersymmetry-breaking B-terms, this direction is flat at tree level. At loop level there are a priori three possibilities. Either the real part of the B-terms and the supersymmetric masses run differently, and then the spectrum along the $S^3$ becomes either gapped or tachyonic, or they run in the same way, preserving the masslessness of the $S^3$ scalars.

Perturbative corrections to supersymmetric gauge theories of the kind we are interested in were investigated by Parkes and West~\cite{Parkes:1982tg,Parkes:1983ib,Parkes:1983nv,Parkes:1984dh}. They considered the addition of mass terms that preserve some supersymmetry, and they also applied the spurion method~\cite{Girardello:1981wz} to study theories in which supersymmetry is completely broken. They derived several powerful all-loop results, a subset of which we now combine for our analysis.

The first step in the breaking of supersymmetry is the addition of a mass term to the $\cN = 4$ superpotential, resulting in an equal-mass $\cN = 1^{\star}$ theory. This theory was shown to remain finite to all loops \cite{Parkes:1982tg}.

The second step is to add to the Lagrangian the particular B-terms induced by the KS background. These B-terms have the ``$X^2-Y^2$'' form \eqref{eq:1and20-20}. It was shown that adding such terms, in conjunction with supersymmetric masses, to $\cN = 4$ Super Yang-Mills, preserves finiteness to all orders in perturbation theory~\cite{Parkes:1983ib}.\footnote{In addition, the ``$X^2-Y^2$'' terms preserve finiteness to all orders when added to a finite $\cN = 2$ theory~\cite{Parkes:1983nv}, and preserve two-loop~\cite{Parkes:1984dh} and one-loop~\cite{Jones:1984cu} finiteness in $\cN = 1$ gauge theories.}

Combining these two results implies that neither the supersymmetric masses nor the B-terms receive any higher-loop correction, and hence the masslessness of the three scalars along the $S^3$ direction is preserved to all orders in the loop expansion. This absence of these perturbative corrections applies not only to the bosonic masses, but to all terms in the Lagrangian of the anti-D3 brane gauge theory.

\section{Physical interpretation}\label{sec:phys}

In this section we discuss the physical significance of the all-loop result obtained above. To do this, we first described the region of parameter space in which we work, and then compare it to previous results. The open string loop expansion is valid for $g_s \overline{N}_{\!3} \ll 1$ and, since the $U(1)$ sector is free, we focus on the $SU(\overline{N}_{\!3})$ sector and hence work at $\overline{N}_{\!3} > 1$.

This is the opposite regime to the one used to analyze fully-backreacted antibranes in supergravity ($g_s \overline{N}_{\!3} \gg 1$) and we believe that the striking agreement between our results and those of \cite{Bena:2014jaa} strongly suggests that the scalars corresponding to motion along the $S^3$ at the bottom of the KS solution remain massless for all values of $g_s \overline{N}_{\!3}$.

\begin{figure}[t!]
\begin{center}
\includegraphics{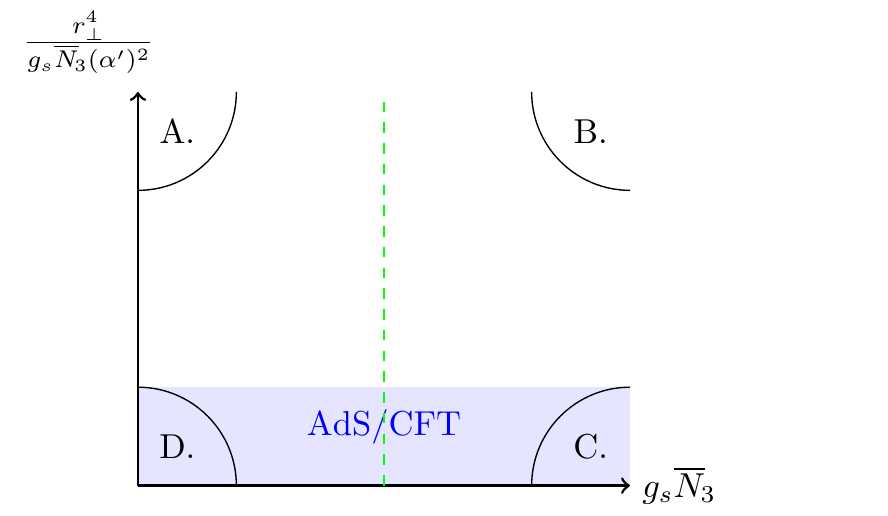}
\end{center}
\caption{
An illustration of the different regimes of parameters. Some recent investigations carried out in the different limits include: A. \cite{Michel:2014lva}, B. \cite{Cohen-Maldonado:2015ssa}, C. \cite{Bena:2014jaa}, D. The present work. The AdS/CFT-like decoupling limit is shaded in blue, and the vertical (green) dashed line is $g_s \overline{N}_{\!3} = 1$.\label{fig:reg}}
\vspace{0.5mm}
\end{figure}

We work in the usual low-curvature supergravity limit, where the length scale, $L$, associated to the curvature of the background is much larger than the string length, $\ell_s/L \ll 1$. This suppresses higher-derivative terms in the brane action, coming from the Taylor expansion of the supergravity fields, such as
\begin{equation}
g_{mn}(\varphi) = \sum_{\alpha=0}^{\infty} \frac{\lambda^\alpha}{\alpha!} \varphi^{n_1} \ldots \varphi^{n_\alpha} \partial_{n_1} \ldots \partial_{n_\alpha} g_{mn}|_{0}\,,
\end{equation}
where $\sqrt{\lambda} \partial_m \sim \ell_s/L$.

Thirdly, we work in the AdS/CFT decoupling regime, in which the low-energy gauge theory on the brane decouples from the supergravity fields. This regime corresponds to sending the distance $r_{\perp}$ from the branes to zero with $r_{\perp}/\alpha'$ fixed~\cite{Maldacena:1997re,Aharony:1999ti}, which means that one has
\begin{equation}\label{eq:asdregime}
\frac{r_{\perp}^4}{g_s \overline{N}_{\!3} (\alpha')^2} \;\ll\; 1\,.
\end{equation}
Since we work in the weakly-coupled gauge theory, the conjectured bulk dual is strongly-curved and the corresponding sigma-model is strongly-coupled. We depict our regime and compare it to the regimes considered in other works in Figure \ref{fig:reg}.

\begin{figure}[h!]
	\begin{center}
		\begin{tabular}{m{2.4cm}m{1cm}m{2.1cm}m{0.8cm}m{.2cm}}
\includegraphics{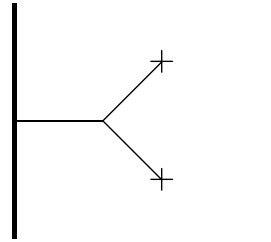}
			&$+$&
\includegraphics{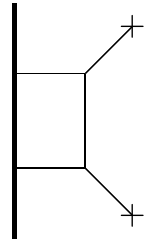}
			& $+$ & $\ldots$
		\end{tabular}
	\end{center}
\caption{Contributions to the antibrane potential within an EFT of the brane and supergravity fields~\cite{Michel:2014lva}. Crosses represent external supergravity fields.\label{fig:bulkdiagram}}
\vspace{8mm}
\end{figure}

\begin{figure}[h!]
\begin{center}
\begin{tabular}{m{3.0cm}m{.4cm}m{2.6cm}m{.4cm}m{3.8cm}m{.4cm}m{.4cm}}
\includegraphics{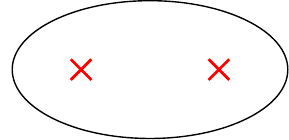}
& $+$ &
\includegraphics{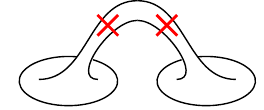}
& $+$ &
\includegraphics{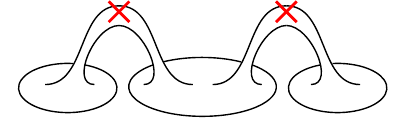}
& $+$ & $\ldots$
\end{tabular}
\end{center}
\caption{Closed-string loop diagrams of which the diagrams in Fig.\;\ref{fig:bulkdiagram} are limits.
Crosses represent closed string vertex operators corresponding to the external legs in Fig.\;\ref{fig:bulkdiagram}. \label{fig:closedstring}
\vspace{3mm}
}
\end{figure}

The difference between our approach and that of \cite{Michel:2014lva} is that the latter considers a low-energy EFT involving both brane and supergravity fields, that is valid for $r_{\perp} \gg \ell_s$.
The diagrams that enter in the calculation of this EFT, depicted in Fig.~\ref{fig:bulkdiagram}, are the massless-closed-string limit of the string diagrams in Fig.~\ref{fig:closedstring}. Our diagrams are simply the light-open-string limit of the same diagrams. For example, if we insert external open-string vertex operators in Fig.\;\ref{fig:closedstring}, we see that this diagram and the one-loop open-string diagrams in Fig.\;\ref{fig:openstring} are the same.
Furthermore, as mentioned above, the would-be field theory corrections to the fermion and boson mass terms can be thought of as representing the correction to the corresponding supergravity fields caused by the backreaction of the brane. Therefore, the diagrams in Fig.~\ref{fig:bulkdiagram} and those in Fig.\;\ref{fig:oneloop} compute the same quantity in two different regimes.

The advantage of our regime is that it allows one to do precise calculations, which, given the current technology of string loop calculations in Ramond-Ramond backgrounds, does not appear possible in the approach of \cite{Michel:2014lva} in the near future. Furthermore, it is entirely possible that if such calculations were done, the exact cancellations leading to the flat directions along the $S^3$ may not survive in the regime of parameters of \cite{Michel:2014lva}, and thus the corresponding EFT might have tachyonic terms. A priori, such terms could arise both for multiple branes and for a single brane.\footnote{Although, of course, for a single brane, such a tachyon cannot be interpreted as indicating brane-brane repulsion.}

If such tachyonic terms are present, the dynamics becomes complicated and falls beyond what can be computed with current technology, either analytically or numerically. One can ask whether there may be a nearby metastable minimum that could be used for uplifting to de Sitter (as discussed in \cite{Michel:2014lva}); such a minimum would first need to be found, and then a stability analysis would need to be performed. An analytic approach would be to use the brane effective action, which involves calculating string diagrams in a Ramond-Ramond background. A numerical approach would be to build a fully-backreacted solution and analyze its stability; this is a cohomogeneity-three problem. Both approaches are currently computationally out of reach. Taken together with the existing results on antibrane instabilities, we believe that the burden of proof for claiming the existence of a new metastable minimum rests on those who would make such a claim. Hence, our reluctance to share the optimism expressed in \cite{Michel:2014lva} regarding anti-D3 brane uplift \cite{Kachru:2003aw}.

In the analysis of fully backreacted antibranes in the KS solution \cite{Bena:2014jaa} it was argued that the cancellation of the bosonic potential along the $S^3$ is not the full story. More specifically, this cancellation comes from a nontrivial relation between the real part of the B-terms and the trace of the boson mass matrix, but in addition the B-terms could also have imaginary parts that are not prohibited by the symmetry of the problem, and one therefore expects to find them generically. Such a term would give rise to tachyonic instabilities in off-diagonal directions~\cite{Bena:2014jaa}.

In our analysis, the imaginary parts of the B-terms are not present at tree level, and are also not generated by loops. However, since there are no symmetries protecting against such terms, it is entirely possible that they will arise non-perturbatively in $g_s \overline{N}_{\!3}$ or at subleading order in the expansions discussed above that take us away from the regime of parameters in which we work (the blue region in Figure \ref{fig:reg}).

If such corrections preserve the balance between the real part of the B-terms and the scalar mass terms, then non-zero imaginary parts of the B-terms would give rise to tachyons. While one expects this balance to be preserved in the decoupling limit when interpolating from weak to strong coupling, there is a priori no reason why it should be preserved away from the decoupling limit. This in turn could generate a gap or could lead to tachyons even without non-zero imaginary parts of the B-terms. Thus, while our result does not prove that the potential has a tachyonic direction, it shows that the potential is vulnerable to tachyonic corrections of the types discussed above, which may be expected to be generically present by standard EFT reasoning.

Another question which one can ask is
whether in our regime of parameters one can see a non-perturbative brane-flux annihilation effect of the type proposed in~\cite{Kachru:2002gs}. If one first considers anti-D3 branes in the S-dual of KS geometry, their worldvolume theory has Higgs vacua. These vacua correspond to the polarization of the anti-D3 branes into D5 branes wrapping an $S^2$ inside the $S^3$ at the bottom of the throat. However, in the limit in which we work,  $\ell_s/L \to 0$, the height of the energy barrier that these D5 brane have to traverse in order to trigger brane-flux annihilation is infinite, and hence the tunneling probability of the anti-D3 branes is zero.
It is almost certain that this height is also infinite in the KS geometry in our limit, because the size of the $S^3$ that the NS5 brane with anti-D3 brane charge has to sweep out diverges.

We note in passing that it has been suggested that anti-D3 brane singularities may possibly be resolved by polarization into NS5-branes \cite{Cohen-Maldonado:2015ssa}. We do not study such polarization in this work, since it requires one to work in the regime of parameters $g_s N \gg 1$, as discussed in the Introduction. We recall however that it was argued in \cite{Bena:2015kia} that a tachyonic term in the polarization potential renders this NS5 configuration unstable to developing shape modes that break spherical symmetry. The brane-brane repulsion results in the tachyonic accumulation of anti-D3 density (encoded by the worldvolume flux on the NS5) near the endpoints of the major axis of an NS5 ellipsoid inside the $S^3$. Another possible manifestation of brane-brane repulsion is the expulsion of anti-D3 branes from the NS5 (analogous to the anti-M$2$ expulsion discussed in \cite{Bena:2014bxa}).

\section{Discussion}\label{sec:discussion}

In this paper we have computed the potential of anti-D3 branes placed at the bottom of the KS throat, in the regime $0 < g_s {N} \ll 1$ and in the AdS/CFT limit, to all orders in perturbation theory.
We first computed the tree-level Lagrangian, and determined the pattern of (soft) supersymmetry breaking.
We then applied certain well-established results on finiteness to show that this Lagrangian does not receive corrections to all loops in perturbation theory, and hence three of the scalars on the worldvolume of the anti-D3 branes in KS remain massless to all orders in the loop expansion.
The fact that this result matches the one obtained in the fully-backreacted regime ($g_s \overline{N}_{\!3} \gg 1$) in Ref.~\cite{Bena:2014jaa} is strong evidence that the spectrum of anti-D3 branes in KS does not become gapped in any regime of parameters where exact calculations can be made.
Furthermore, since there is no symmetry prohibiting an imaginary B-term in the effective action on the branes, and since such a term will always introduce tachyons, the optimism about brane-brane-repelling tachyons disappearing when $g_s \overline{N}_{\!3} \ll 1$ appears premature.

Although the explicit analysis performed in this paper is for the KS background, we expect the masslessness of some of the worldvolume scalars to be a generic feature of anti-D3 branes in conical highly-warped geometries: For a conical geometry to be regular at its bottom it is necessary to have some finite cycle, such as the $S^3$ of the deformed conifold or the $S^2$ of the resolved conifold.
From the perspective of the worldvolume theory of an antibrane at the bottom of this geometry, this would imply that some of the scalars are massless at tree level and hence the B-term will be nonzero.

In a generic conical geometry with ISD fluxes the theory on the anti-D3 branes will contain fermion bilinears and scalar trilinears, that will consist of both ``supersymmetric'' terms (of the kind we found in KS) coming from the primitive (2,1) components of $G_3$, and also non-supersymmetric terms coming from the (0,3) and the non-primitive (1,2) $G_3$ components. For example, a (0,3) component would introduce a $h_{ijk} \Tr\{\phi^i\phi^j\phi^k\} + \textrm{h.c.}$ scalar trilinear term and a gaugino mass $M \lambda \lambda$. The relation between $h_{ijk}$ and $M$ is exactly of the same form as that between the supersymmetric fermion masses and scalar trilinear couplings that we analyzed in detail in Section \ref{ssec:superpot}, and this, combined with the vanishing of the mass supertrace at tree level~\cite{Bena:2015qfa}, implies that the beta-functions of the theory will vanish both at one and two loops \cite{Martin:1993zk,Yamada:1994id,Jack:1994kd,Jack:1999ud}.\footnote{In particular, this implies that the mass supertrace will remain zero at one and two loops~\cite{Bena:2015qfa}.} Hence, the scalars that are massless at tree level will remain massless at least to two loops.

A similar argument can be made for a more generic background that contains a combination of primitive (2,1), (0,3), and non-primitive (1,2) flux: these fluxes would give rise to a symmetric $4\times 4$ fermion mass matrix that can be diagonalized (this corresponds to changing the complex structure), and the resulting theory would be the one with (2,1) and (0,3) fluxes discussed above.

Our result appears to be in tension with the argument that the spectrum of anti-D$3$ branes in the KS geometry is gapped \cite{Kachru-talk}, and also with the related argument that separated anti-D3 branes at the bottom of the KS solution should be screened by flux and therefore should attract each other~\cite{DeWolfe:2004qx}.
To explicitly compute the effect of this screening on the potential between two anti-D3 branes, one must perform a full string calculation, which cannot be done with current technology. However, in the limit where the branes are close to each other, this string calculation reduces to our field-theory calculation, which finds  that the tree-level masslessness of three of the six scalars that describe the anti-D$3$ branes is preserved to all loops. Therefore, the spectrum of the anti-D$3$ branes remains un-gapped in our regime of parameters. Hence, the intuition that  antibranes at the bottom of KS are screened and therefore attract each other does not appear to give the correct physics in either of the two regimes of parameters where precise calculations can be done: the regime we have considered and the large-backreaction regime \cite{Bena:2014jaa}.

The absence of a gap may also be problematic for phenomenological applications. When the KS throat is glued to a compact manifold, the gluing introduces perturbations to the KS throat, which in turn can give very small masses to the scalars that are massless in the infinite KS solution. These masses were estimated in Ref.~\cite{Aharony:2005ez}, and were found to be exponentially smaller than the typical mass scale of the light fields at the bottom of the throat, and hence phenomenologically problematic. Had our calculations found instead that the inter-brane degrees of freedom were gapped, there would have only been three such light fields, corresponding to the center-of-mass degrees of freedom, and these fields could presumably have been uplifted in some other way. However, uplifting $3\overline{N}_{\!3}$ massless modes appears a more and more onerous task as one increases $\overline{N}_{\!3}$.
It would be interesting to compare these corrections to those discussed in Section \ref{sec:phys}, in particular to see whether any possible tachyonic term could overwhelm these very light masses.


We have also discussed the regime of parameters in which our calculation is done and its relation to the brane effective action approach of \cite{Michel:2014lva}: The Feynman diagrams that enter in our field-theory calculation arise from the light-open-string limit of string diagrams, which in the opposite massless-closed-string limit, reduce to the supergravity amplitudes considered in~\cite{Michel:2014lva}.

There is another difference between these approaches, which has to do with the number of anti-D$3$ branes we consider. The theory on $\overline{N}_{\!3}$ anti-D$3$ branes has a $U(\overline{N}_{\!3})$ gauge group, and its dynamics can be split in an $SU(\overline{N}_{\!3})$ sector and a $U(1)$ sector. The $U(1)$ sector describes the center-of-mass motion of the branes, and is a free theory, reflecting the fact that there is no potential to the stack of antibranes (or a single antibrane) moving together on the $S^3$, as mentioned in~\cite{Michel:2014lva}. It is important to note that this fact is not a result of a calculation done using the brane effective action, but simply a result of symmetry considerations.\footnote{There has also been interest in studying bound states of an anti-D3 brane and an O$3^-$ plane~\cite{Uranga:1999ib,Kallosh:2014wsa,Bergshoeff:2015jxa,Kallosh:2015nia,Garcia-Etxebarria:2015lif}, which entirely removes the six anti-D3 scalar degrees of freedom, and hence any potential brane-brane repelling tachyons.
Besides the fact that these objects have finite charge but zero mass and so strongly violate the BPS bound, these constructions so far lack the explicitness available in KS.
}

As we have noted, when going away from our field-theory limit, both the $U(1)$ and the $SU(\overline{N}_{\!3})$ sectors may receive corrections, which are capable of introducing a gap or a tachyon. There is no symmetry that prohibits these corrections, even for the $U(1)$ sector, although the interpretation of the possible existence of a tachyon for a single antibrane is unclear.



Our result that the brane-brane potential along the $S^3$ at the bottom of the KS solution remains flat agrees exactly with the strong-coupling calculation of \cite{Bena:2014jaa}. In \cite{Bena:2014jaa} it was furthermore argued that the symmetries of the problem do not prohibit the existence of another term in the brane-brane potential (an imaginary part of the B-terms) and thus, following the usual EFT reasoning, one expects that such a term will generically be present. This term does not affect the value of this potential along the $S^3$ but on the other hand gives rise to a brane-brane repelling tachyon along a direction misaligned with the $S^3$.

It is important to ask whether a similar term could possibly also appear and give rise to a brane-brane-repelling tachyon in the weak-coupling regime in which we work. In the field theory on the branes this term is zero at tree level, and one can also show that the beta-functions associated to its running are exactly zero. Hence one possibility is that this term is exactly zero in the weak-coupling regime, and only appears in the large-backreaction regime. However, if one examines the problem a bit deeper, and excludes mathematical oddities, this possibility appears quite unlikely. Rather, if the brane-brane-potential on the $S^3$ remains flat all the way from weak to strong coupling, and the tachyon-inducing term is considerable at large $g_s \overline{N}_{\!3}$, one expects that a leftover of this term, however small, will be visible at weak coupling, perhaps as a non-perturbative effect.
Whenever this term is not exactly zero, a brane-brane-repelling tachyon is present~\cite{Bena:2014jaa}, and thus we expect that this tachyon will be present at all finite values of $g_s \overline{N}_{\!3}$.

Finally, let us comment on the implications of our result for the possibility of using anti-D3 branes in long warped KS-like throats to uplift the cosmological constant and obtain a landscape of metastable de Sitter solutions in String Theory. Our computation found a flat direction in the brane-brane potential which is preserved at all loops, indicating that the system is ungapped and is vulnerable to tachyons which, from an EFT perspective, are likely to be present.
Therefore anti-D3 brane uplifting mechanisms remain questionable even when backreaction is small.

Our paper thus contains yet another calculation that a priori could have either agreed or disagreed with the viability of anti-D3 brane uplifting constructions.
Taken together with the negative results obtained in the large backreaction regime~\cite{Bena:2009xk,Bena:2010gs,Dymarsky:2011pm,Bena:2011wh,Blaback:2011pn,Bena:2012bk,Gautason:2013zw,Giecold:2013pza,Bena:2012tx,Bena:2012vz,Bena:2014bxa,Bena:2012ek,Bena:2013hr,Buchel:2013dla,Blaback:2014tfa},
our result further adds to the evidence against the existence of a de Sitter multiverse obtained using antibranes.\footnote{If one desires, one can even quantify this evidence in a Bayesian approach~\cite{Polchinski:2016xto}.}

\section*{Acknowledgments}

We thank Daniel Baumann, Micha Berkooz, Ulf Danielsson, Emilian Dudas, Anatoly Dymarsky, Mariana Gra{\~n}a, Stanislav Kuperstein, Emil Martinec, Stefano Massai, Praxitelis Ntokos, Giulio Pasini, Joe Polchinski, Andrea Puhm, Rodolfo Russo, and Thomas Van Riet for useful discussions. This work was supported by the John Templeton Foundation Grant 48222. The work of I.B. was also supported by the ERC Starting Grant 240210 String-QCD-BH, and by a grant from the Foundational Questions Institute (FQXi) Fund, a donor advised fund of the Silicon Valley Community Foundation on the basis of proposal FQXi-RFP3-1321 (this grant was administered by Theiss Research). The work of J.B. and D.T. was also supported by the CEA Eurotalents program.

\appendix

\section{Conventions}\label{app:conv}

In this appendix we record our conventions and their relation to those used in a selection of related literature~\cite{Grana:2002tu,Myers:1999ps,Grana:2003ek,Camara:2003ku,McGuirk:2012sb,Bergshoeff:2015jxa}. Our conventions are:
\begin{itemize}
\item $G_3 \equiv F_3 - e^{-\phi}H_3$ and $E_{MN} \equiv g_{MN} - B_{MN}$, which implies that the RR fields strengths are $F = \d C + H \w C$;
\item Our Hodge-star conventions are those described in Appendix A of \cite{Danielsson:2009ff};
\item The anti-D$3$ brane worldvolume theory has interaction terms induced by ISD fluxes, $\star_6 G_3 = i G_3$.
\end{itemize}
Our  $H_3 = \d B_2$ has the opposite sign compared to that of Refs.~\cite{Myers:1999ps,Camara:2003ku,McGuirk:2012sb,Bergshoeff:2015jxa}. Ref.~\cite{Grana:2003ek} does not follow the same Hodge-star conventions as us, and Ref.~\cite{Camara:2003ku} switches to a mostly-minus signature for their four-dimensional theory while we keep strictly to a mostly-plus signature. Note that we also start from the string-frame brane action, which is a choice that becomes irrelevant after the constant rescaling done in Eq.~(\ref{eq:rescale}).

In addition, we differ from Ref.~\cite{Myers:1999ps} in our conventions for the RR fields. We have $F_n \sim g_s^{-1} H_3$ while Ref.~\cite{Myers:1999ps} has $F_n \sim H_3$.
Finally, in our conventions we have $\ell_s^2 = \alpha'$.

\section{Fermion masses from D5 polarization}\label{sec:D5}

As discussed in Section \ref{ssec:superpot}, the relative normalizations of the bosonic and fermionic actions can be directly derived from the D$5$ polarization potential. In the KS background a D$5$ brane carrying anti-D$3$ brane charge and wrapping the shrinking $S^2$ at a finite distance, $\tau$, from the bottom of the deformed conifold, has the action
\begin{equation}
V_{\textrm{D}5} = 2\pi \bar{N}_3 c_2 \tau^2 - c_3 \tau^3 + \frac{1}{2\pi \bar{N}_3} c_4 \tau^4\,
\end{equation}
where $c_{2,3,4}$ are constants.
Details of the derivation of this potential can be found for example in Ref.~\cite{Bena:2012vz}. This potential has no minimum away from $\tau = 0$, and hence anti-D3 branes in KS cannot polarize into D5 branes.

We are interested in the coefficient $c_2$: this is proportional to $\partial_\tau^2( e^{4A}+ \alpha)|_{0}$ and is hence proportional to the mass-squared of the scalars that
correspond to motion away from the bottom of the warped deformed conifold.

 Now, if we deform the potential by taking $c_2 \to c_2/2$, the potential can now be written as a perfect square. Explicitly, the deformed potential is
\begin{equation}
\tilde{V}_{\textrm{D}5} = \frac{b^2}{128} \frac{g_s}{\pi \bar{N}_3} \tau^2 \left( \tau - \frac{2\sqrt{2}}{b^2} e^{4A_0} |G_3| \frac{\pi \bar{N}_3}{g_s}  \right)^2\,.
\end{equation}
This expression has been translated into our conventions using the local $\mathbb{R}^6$ coordinates of Ref.~\cite{Bena:2014jaa} (modified to be consistent with our complex coordinates defined in Eq.~(\ref{eq:cplxcoord})) and the KS conventions found in Ref.~\cite{Bena:2012vz}.

This deformed potential obtained by $c_2 \to c_2/2$ has a supersymmetric minimum at non-zero $\tau$. Thus, the mass-squared of the three massive Hermitian scalars in the undeformed potential is twice its would-be supersymmetric value, i.e.~the mass-squared of the fermions.



\newpage

\bibliography{refs}

\bibliographystyle{utphysmodb}

\end{document}